\documentclass[conference]{IEEEtran}
\usepackage{cite}
\usepackage{amsmath,amssymb,amsfonts}
\usepackage{algorithmic}
\usepackage{graphicx}
\usepackage{textcomp}
\usepackage{xcolor}
\usepackage{fancyhdr}
\usepackage[hyphens]{url}
\usepackage{caption}
\usepackage{xfrac}
\def\BibTeX{{\rm B\kern-.05em{\sc i\kern-.025em b}\kern-.08em
    T\kern-.1667em\lower.7ex\hbox{E}\kern-.125emX}}

\title{Energy-efficient Dense DNN Acceleration with Signed Bit-slice Architecture}

\author{\IEEEauthorblockN{Dongseok Im,
Gwangtae Park,
Zhiyong Li, 
Junha Ryu, and
Hoi-Jun Yoo}
\IEEEauthorblockA{School of Electrical Engineering}
\IEEEauthorblockA{Korea Advanced Institute of Science and Technology (KAIST)}
\IEEEauthorblockA{Daejeon, Republic of Korea}
\IEEEauthorblockA{Email: dsim@kaist.ac.kr}
}

\begin{document}
\maketitle
\thispagestyle{plain}
\pagestyle{plain}

%%%%%% -- PAPER CONTENT STARTS-- %%%%%%%%

\begin{abstract}
As the number of deep neural networks (DNNs) to be executed on a mobile system-on-chip (SoC) increases, the mobile SoC suffers from the real-time DNN acceleration within its limited hardware resources and power budget.
Although the previous mobile neural processing units (NPUs) take advantages of low-bit computing and exploitation of the sparsity, it is incapable of accelerating high-precision and dense DNNs.
This paper proposes energy-efficient signed bit-slice architecture which accelerates both high-precision and dense DNNs by exploiting a large number of zero values of signed bit-slices.
Proposed signed bit-slice representation (SBR) changes signed $1111_{2}$ bit-slice to $0000_{2}$ by borrowing a $1$ value from its lower order of bit-slice. 
As a result, it generates a large number of zero bit-slices even in dense DNNs.
Moreover, it balances the positive and negative values of 2's complement data, allowing bit-slice based output speculation which pre-computes high order of bit-slices and skips the remaining dense low order of bit-slices.
The signed bit-slice architecture compresses and skips the zero input signed bit-slices, and its zero skipping unit also supports the output skipping by masking the speculated inputs as zero. 
Additionally, the heterogeneous network-on-chip (NoC) benefits exploitation of data reusability and reduction of transmission bandwidth.
The paper introduces a specialized instruction set architecture (ISA) and a hierarchical instruction decoder for the control of the signed bit-slice architecture.
Finally, the signed bit-slice architecture outperforms the previous bit-slice accelerator, Bit-fusion, over $\times3.65$ higher area-efficiency, $\times3.88$ higher energy-efficiency, and $\times5.35$ higher throughput.

\end{abstract}

\section{Introduction}
Recently, many mobile devices have realized 2-D and 3-D vision tasks using deep neural network (DNN) such as photography improvement~\cite{bokeh}, visual question and answering~\cite{embodiedQA}, and AR/VR~\cite{facebook}.
A mobile AP in a mobile device~\cite{SamsungNPU, mediatek, hololens2, facebook}, which consists of a CPU, a GPU, and application IPs, performs the various functions in a single chip.
Then, the mobile AP integrates a neural processing unit (NPU) to accelerate the DNNs in a real-time.
However, as increasing the number of DNNs to be implemented even in a single vision task~\cite{imvotenet ,embodiedQA}, the mobile NPU is suffered from accelerating the complex multiple DNNs within limited memory bandwith, computing resources, and restricted power budget.

Lowering the bit precision of DNNs can alleviate the deficiencies of a mobile NPU with minimal accuracy degradation. 
It lessens the memory bandwidth and on-chip memory footprint due to low bit-width of data, and it increases the computing throughput by integrating a large number of low bit multiplier-and-accumulate (MAC) units.
Therefore, bit-slice accelerators~\cite{bitfusion, bitblade, hnpu, SamsungNPU, mediatek} accelerate various bit-precision of DNNs with numerous number of low bit MAC units by dynamically matching the bit-width in a spatial- and time-multiplexing method.
However, accuracy-sensitive tasks such as image super-resolution~\cite{fsrcnn} and monocular depth estimation~\cite{monodepth2} require higher precision than object classification~\cite{pointnet++, dgcnn} and object detection~\cite{yolov3, votenet} tasks.
Since the inference time of bit-slice architecture is increased linearly as the precision of the data, high bit-precision DNN acceleration deteriorates the performance of bit-slice architecture.

Exploiting the zero sparsity of DNNs can increase a hardware performance by skipping redundant computations caused by zero values. 
For example, the ReLU activation function generates numerous zero values in input data, and the zero skipping unit skips the computations of them with minimum of latency.
Furthermore, after decomposing the 2's complement of fixed-point data to bit-slices, additional sparsity is occurred in a small positive value of data whose high order of bit-slices are zeros.
Therefore, zero bit-slice skipping architecture~\cite{hnpu} takes the advantages of both low bit computing and zero bit-slice skipping, which increases the energy-efficiency even in high bit-precision DNNs. 

\begin{figure} [!t]
\includegraphics[width=\columnwidth]{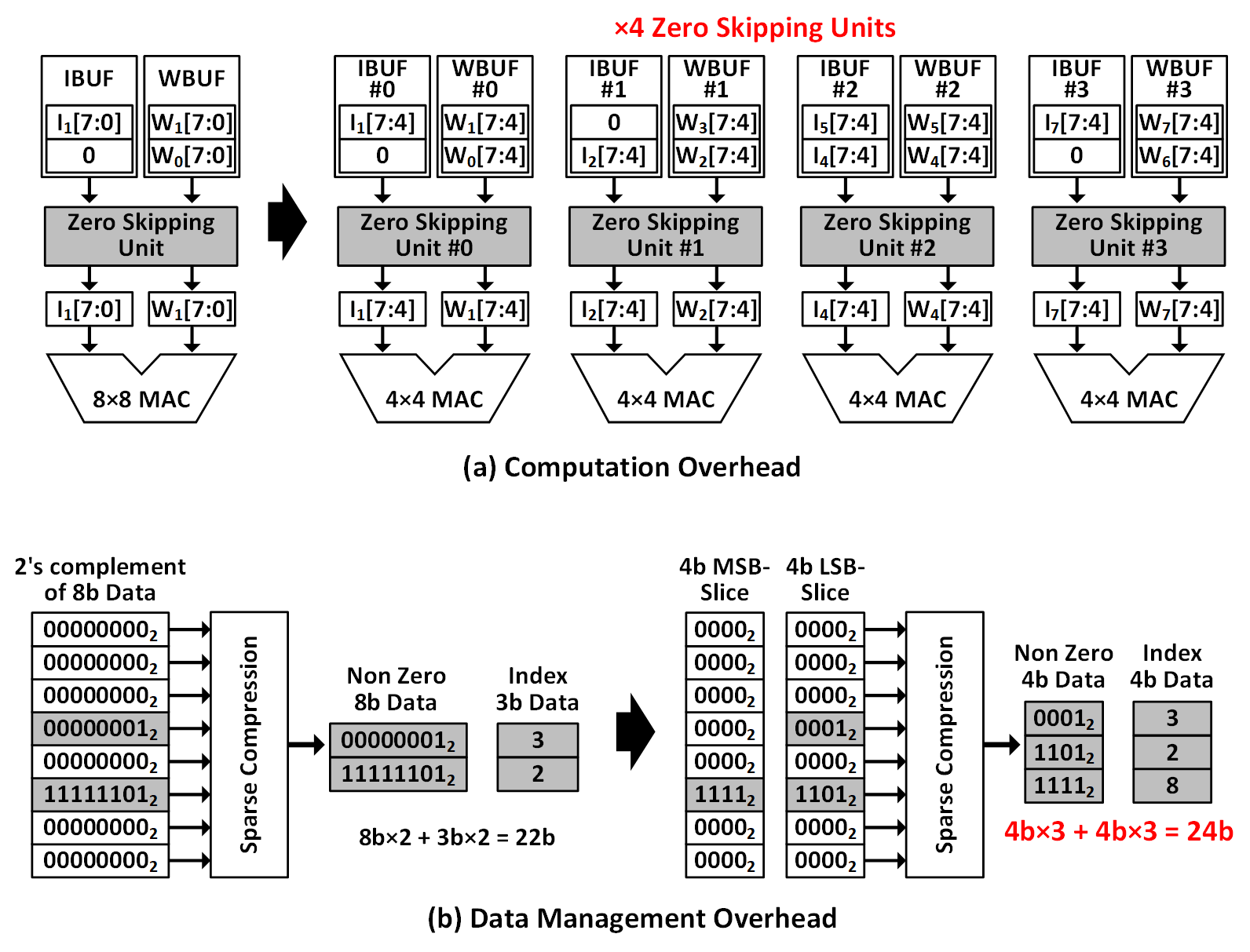} 
\centering
\caption{Hardware challenges in bit-slice skipping architecture in terms of (a) computation and (b) data management.}
\label{fig:fig1}
\end{figure}

Zero skipping method requires a zero skipping unit which pre-fetches the non-zero inputs and loads the corresponding weights without any latency.
However, as shown in Fig.~\ref{fig:fig1}, since 4-bit bit-slice architecture adopts $\times$4 numbers of MAC units compared to 8-bit fixed bit-width MAC units, it requires $\times$4 numbers of zero skipping units, which increases the more area and power consumption. 
In the data management, the sparse data compression reduces the number of data transactions and memory footprint by encoding the sparse data to non-zero data and its index.
However, the bit-width of non-zero index becomes a large relative to the non-zero data after decomposing the data which halves the bit-width and doubles the numbers.
As a result, the compression ratio is easily lower than the fixed bit-width data.

\begin{figure} [!t]
\includegraphics[width=\columnwidth]{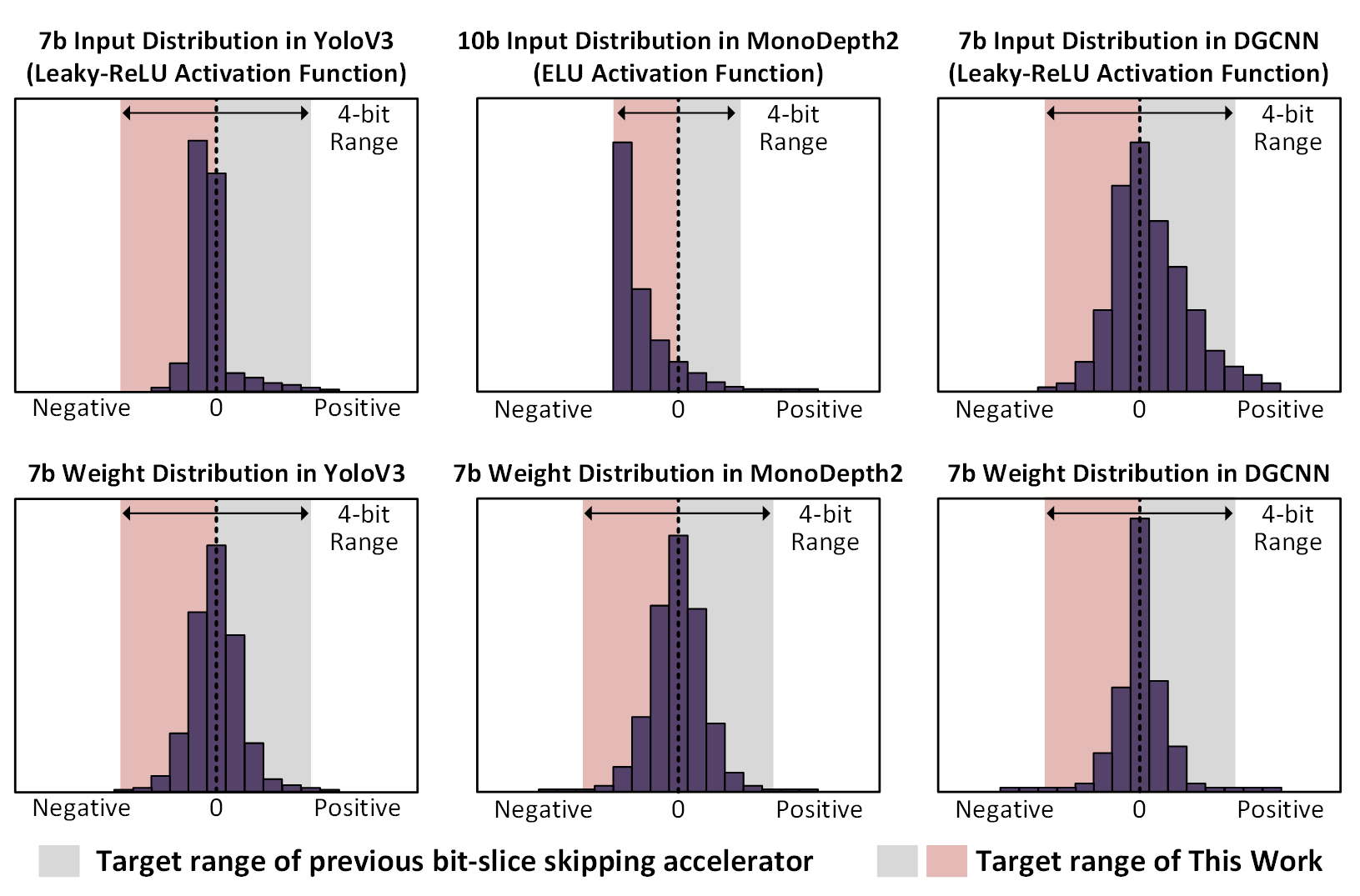} 
\centering
\caption{Input and weight distribution in DNNs with a target range of the previous zero bit-slice skipping accelerator and the signed bit-slice architecture.}
\label{fig:fig3}
\end{figure}

To make matters worse, various state-of-the art DNN use non-ReLU activation functions such as Leaky ReLU or ELU~\cite{elu}, so the input sparsity declines precipitously.
In the case of weight data, since weight data follows the gaussian distribution during DNN training~\cite{lecun2012efficient, glorot2010understanding, he2015delving}, it presents the dense weight data.
Therefore, the zero bit-slice skipping architecture cannot exploit a lot of zero data anymore as shown in Fig.~\ref{fig:fig3}.
Furthermore, the additional zero bit-slices are only obtained at the small positive value because the small negative data of 2's complement number produces $1111_{2}$ bit-slices.
As the amount of zero bit-slices are not enough in a non-ReLU activation function and weights data, it worsens the overheads of the zero bit-slice skipping architecture. 

\begin{figure} [t]
\includegraphics[width=\columnwidth]{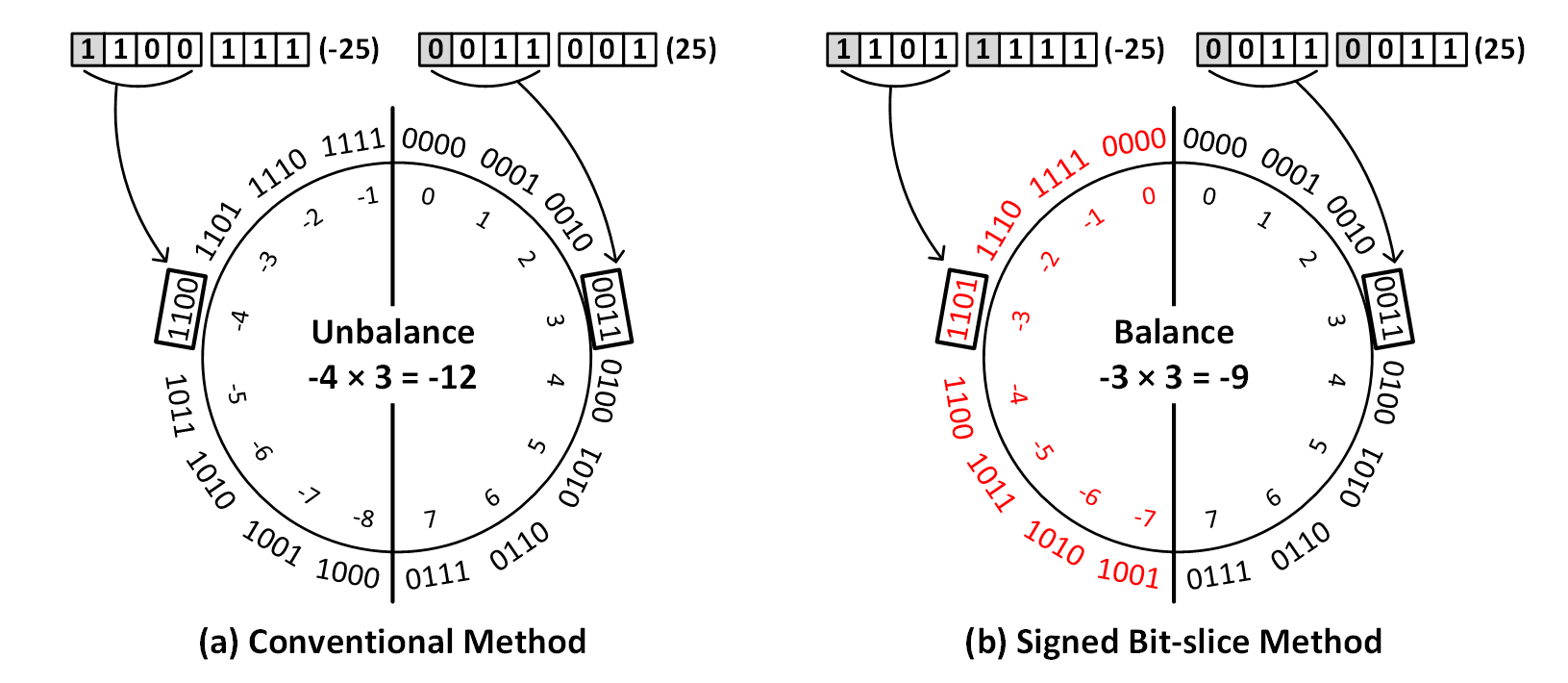} 
\centering
\caption{(a) Unbalance in bit-slice of positive and negative numbers, and (b) balance in signed bit-slice.}
\label{fig:fig4}
\end{figure}

Instead of exploitation of the sparse input and weight data, the output skipping architecture~\cite{song2018prediction, im20204} exploits the output sparsity which is presented by a max pooling operation. 
Especially, 3-D point cloud based DNNs~\cite{pointnet++,votenet,imvotenet,dgcnn} exploit the large scale (e.g. 64-to-1) max pooling operation to extract the maximal feature, so a large number of convolution operations can be removed by exploiting the output sparsity.
Therefore, the output skipping architecture speculates these redundant computations by pre-computing with high order of bit-slice and skips them to achieve high energy-efficiency.
However, the speculation is usually failed because of unbalance of 2's complement number between positive and negative in Fig.~\ref{fig:fig4}.
For example, high order of bit-slice of $1100111_{2}$ $(-25)$ and $0011001_{2}$ $(25)$ is $1100_{2}$ $(-4)$ and $0011_{3}$ $(3)$ respectively by using conventional bit-slice decomposition.
Then, the speculation output of $(-25) \times (-25)$ is 16, but the speculation output of $(25) \times (25)$ is 9.
Their speculation outputs are different even the actual multiplication results are same, which increases the speculation errors.
Therefore, the previous works cannot reduce the bit-width of bit-slices to be used for the speculation because of big speculation errors, and they have to exploit the large bit-width of bit-slices for the speculation which limits the hardware performance enhancement.

The granularity of workload allocation and instruction fetch of the bit-slice architecture increases by controlling a large number of low bit MAC units. 
Therefore, the routing complexity is increased to transfer the data to all of MAC units flexibly.
Moreover, the number of issuing instruction is increased to control all of MAC units that the CPU becomes more busy. 
As a result, the flexible bit-slice architecture with its specialized instruction set architecture (ISA) are necessary in a mobile SoC.

To solve the challenges of bit-slice architecture, the paper presents a new method of bit-slice decomposition which produces sparse signed bit-slices, and the energy-efficient signed bit-slice architecture supports signed bit-slice based in-out skipping with high data compression ratio even in the dense DNN caused by non-ReLU activation functions.
Furthermore, the heterogeneous network-on-chip (NoC) can transfer both input, weight, and output data and partial sum data efficiently.
Finally, a specialized ISA enables the software optimization of the signed bit-slice architecture. 
The summary of the signed bit-slice architecture is below:
\begin{itemize}
\item A new binary representation of 2's complement data, signed bit-slice representation (SBR), increases the sparsity of bit-slices not relying on a ReLU activation function.
SBR adds a sign bit to each bit-slice, and it makes signed $1111_{2}$ bit-slices to $0000_{2}$ by borrowing a 1 value from their lower order of bit-slice.
Therefore, the sparsity of bit-slices becomes high even in non-ReLU activation function, allowing the high performance enhancement and high data compression ratio.
Furthermore, it balances the positive and negative values of 2's complement data which can exploit the output speculation by using the low bit of the data.
\item A signed MAC unit is designed for the high efficiency.
The previous bit-slice architecture has to support both signed and unsigned bit-slices, so the previous MAC unit requires the sign extension unit and sign extended bit-width of a multiplier.
On the other hand, the signed bit-slice architecture integrates the signed MAC unit for the signed bit-slices which does not require the sign extension unit and an enlarged MAC unit.
Therefore, the signed MAC unit can improve a MAC efficiency compared to the previous bit-slice architecture.
\item A zero skipping unit is designed for input and output skipping.
Since the SBR generates a large number of zero bit-slices, the zero skipping unit skips the four spatial adjacent input bit-slices if they are all zeros.
Additionally, the zero skipping unit is compatible with the output skipping by masking the speculated input bit-slices to zeros.
As a result, the signed bit-slice architecture increases the hardware performance by exploiting both input and output skipping. 
\item The signed bit-slice architecture supports hybrid zero skipping and hybrid zero compression which depends on the sparsity ratio.
Since the sparsity of data determines the hardware performance, a dynamic sparsity monitoring (DSM) unit monitors the sparsity ratio of input and weight signed bit-slices and the signed bit-slice architecture skips much higher sparse data.
Similarly, the DSM unit determines the zero compression operation by an amount of the sparsity to increase the overall compression ratio.
\item Heterogeneous NoC is adopted for flexible data transaction. Bi-directional 2D mesh based NoC (Bi-NoC) supports the versatile workload allocation by transferring the inputs, weights, and convolution outputs between the cores with exploitation of data reusability.
On the other hand, uni-directional NoC (Uni-NoC) is designed for the accumulation of the output partial sums with optimized bandwidth.
\item Programmable signed bit-slice architecture executes the DNN tasks with its specialized instruction set architecture (ISA). The hierarchical instruction decoder reduces the number of instruction fetches and minimizes the involvement of the CPU.
\end{itemize}

\section{Related Works}
\subsection{Bit-precision Reconfigurable Architecture}
Quantization is the popular optimization method by reducing the bit-precision of the inputs and weights from the full bit data.
Since lowering the bit-precision alleviates the data transactions and computational complexity, many NPUs take advantage of quantization.
Stripes~\cite{stripes} and UNPU~\cite{unpu} use bit-serial computing units to provide reconfigurable bit-precision.
They accelerate the low-bit DNNs by temporally bit-wise computing.
On the other hand, bit-slice architecture composes of bit-slice processing elements (PEs) which use 2-bit, 4-bit, or 8-bit MAC units and dynamically matches the various bit-width of the DNNs in a spatial and temporal execution. 
It shows much higher area-efficiency and energy-efficiency than bit-serial architecture while taking an advantage of bit-flexibility~\cite{bitfusion}.
With this advantage, bit-slice architecture is widely used in the commercial mobile SoC~\cite{SamsungNPU,mediatek}. 
However, bit-serial and bit-slice architecture takes many cycles to implement high bit-precision applications because of time-multiplexed computation.

\subsection{Sparsity Exploiting Architecture}
Sparsity exploitation can enhance the throughput and energy-efficiency of the accelerators.
It skips the redundant computation caused by the zero input or weight data.
Then, the performance enhancement depends on the sparsity of the data.
Pruning method~\cite{pruning} increases the sparsity of the weight data, and weight skipping accelerators~\cite{eie, zhang2016cambricon} boosts up the performance using the pruned weight.
Although pruning method generates zero values in weight data, it requires additional optimization stage such as retraining.
Therefore, the mobile NPU exploits input zero sparsity rather than the weight because of generality~\cite{SamsungNPU}. 
However, input sparsity is presented to the ReLU activation function, zero input skipping accelerators~\cite{albericio2016cnvlutin,SamsungNPU,mediatek} cannot increase the throughput and energy-efficiency at dense DNNs which use the non-ReLU activation functions.

Zero bit-slice skipping architecture~\cite{hnpu} computes the bit-slice data and skips the zero input bit-slices.
Therefore, it can skip more zero computations by skipping the additional redundant computation caused by zero bit-slices.
However, the number of zero bit-slices is limited to the positive values due to 2's complement data, and it cannot exploit the benefits of both zero bit-slice compression and zero bit-slice skipping in dense DNN computation.

\subsection{Output Skipping Architecture}
Output skipping architecture speculates the zero outputs caused by the ReLU activation function and the max pooling layer, and it skips the redundant computations corresponding to the zero outputs.
Previous work~\cite{song2018prediction} speculates the outputs of a small scale (4-to-1) max pooling layer by pre-computing the high order of bit-slices and accumulates them with the remaining low order of bit-slices.
Other work~\cite{im20204} uses the binary convolution of 1-bit weights for the large scale (64-to-1) max pooling layer prediction.
The PEs only compute the predicted maximal outputs using the heterogeneous architecture.
However, the previous works easily fail to speculate the outputs using low bit-width of both input and weight bit-slices due to its asymmetry, which degrades the hardware performance.

\begin{figure} [!t]
\includegraphics[width=\columnwidth]{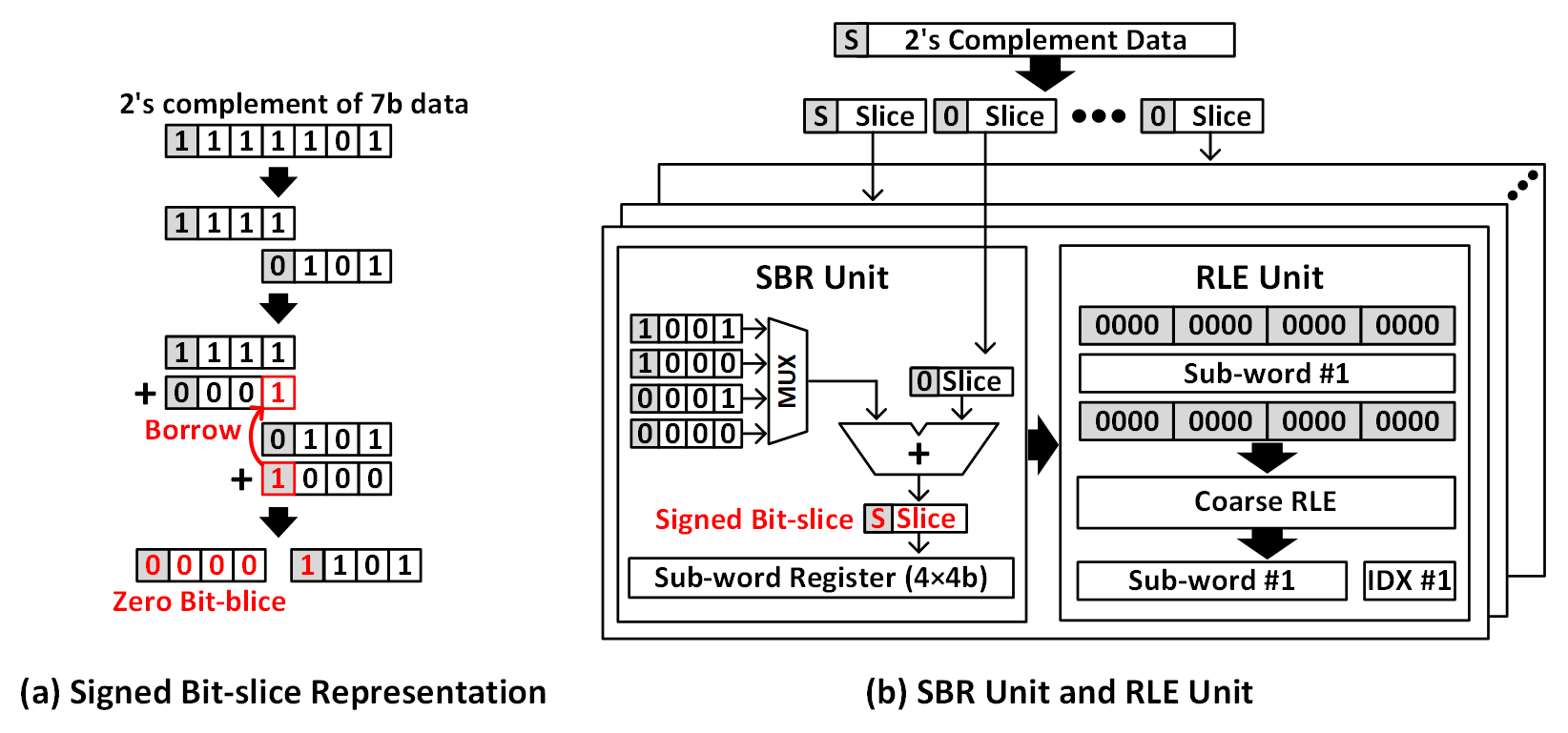} 
\centering
\caption{Concept of signed bit-slice: (a) signed bit-slice representation (SBR) and (b) SBR unit and run-length encoding (RLE) unit.}
\label{fig:fig5}
\end{figure}

\begin{figure*}[!t]
\includegraphics[width=\textwidth]{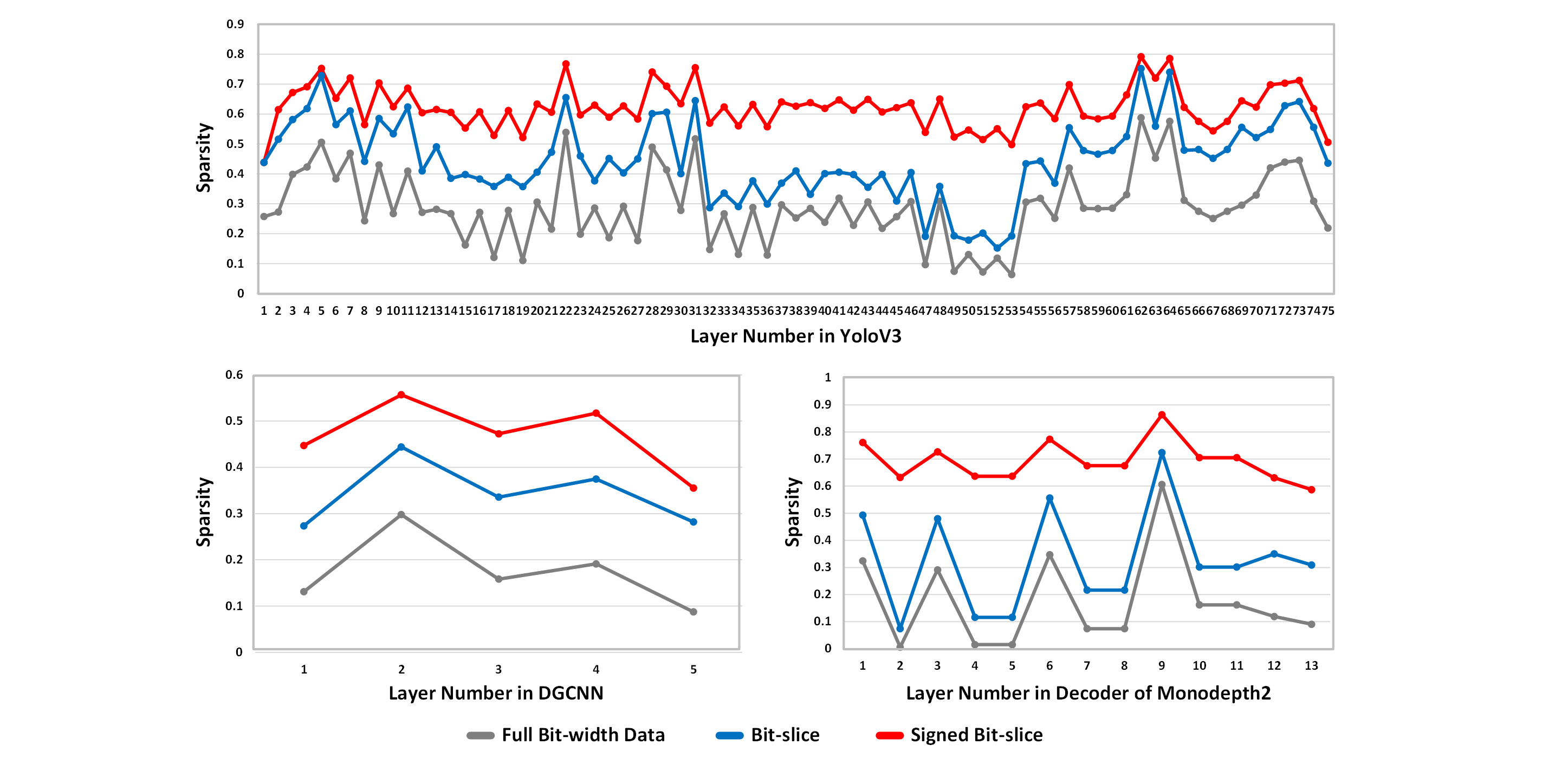} 
\caption{Sparsity ratio of full bit-width data, bit-slice, and signed bit-slice in 7-bit YoloV3, 7-bit DGCNN, and 10-bit decoder of Monodepth2.}
\label{fig:fig9}
\end{figure*}

\section{Signed Bit-slice Architecture}
\subsection{Signed Bit-slice Representation and Its Encoding Unit}
Conventional bit-slice representation~\cite{bitfusion, hnpu} decomposes 2's complement fixed-point data to an MSB bit-slice which is a signed bit-slice and the lower slices which are unsigned bit-slices.
In this work, the SBR adds the sign bit to each unsigned bit-slice to produce the signed bit-slice, and it adds $1$ value by borrowing from its lower bit slice if the data is negative value as shown in Fig.~\ref{fig:fig5}(a). 
For example, $1111101_{2}$ of 2's complement data is decomposed to $1111_{2}$ of signed bit-slice and $101_{2}$ of unsigned bit-slice. 
Then, the SBR makes $101_{2}$ to $0101_{2}$ and adds $1$ to $1111_{2}$ by borrowing from $0101_{2}$.
Finally, they become $0000_{2}$ and $1101_{2}$.
As a result, the SBR makes the majority $1111_{2}$ bit-slices at the small negative values to $0000_{2}$ bit-slices, and the sparsity of signed bit-slice significantly increases.

Fig.~\ref{fig:fig9} describes the sparsity enhancement by the SBR in dense DNNs.
By using the SBR, a high order of signed bit-slice becomes extremely sparse data ($80{\sim}99\%$).
Then, the SBR increases the total signed bit-slice sparsity of YoloV3 by $\times2.14$ and $\times1.39$ higher than the full bit-width data and conventional bit-slice decomposition method, respectively.
Similarly, the sparsity of DGCNN shows $\times2.14$ and $\times1.39$ higher in average.
In the case of dense decoder network of Monodepth2, it has $\times3.94$ and $\times2.11$ higher bit-slice sparsity.
This sparsity enhancement has more potential to boost up the throughput and energy-efficiency with alleviation of the data transactions.

The SBR balances the positive and negative values of the 2's complement data.
It makes both high order of bit-slice of $1111101_{2}$ (-3) and $0000011_{2}$ (3) to $0000_{2}$ (0), 
Therefore, it minimizes the speculation errors while using both low bit-width of input and weight signed bit-slices for the speculation.
For example, the success rate of the output speculation with both 4-bit input and weight signed bit-slices shows roughly $95\%$ in VoteNet, but the previous approach requires 4-bit input bit-slices and 8-bit weight bit-slices to obtain the similar success rate.
Then, the paper adopts the previous speculation approach ~\cite{song2018prediction} which pre-computes the high order of bit-slices and accumulates them with remaining low order of bit-slices.
As a result, the SBR takes advantage of accurate speculation with low bit-width of bit-slices and achieves the high throughput and energy-efficiency enhancement.

Fig.~\ref{fig:fig5}(b) illustrates the SBR unit and the run-length encoding (RLE) unit. 
The SBR unit performs the SBR to the data before being processed at the maxtrix processing (MPU) cores in the signed bit-slice architecture.
It receives the bit-width of the fixed-point data and chooses the value to be added by considering the order of bit-slice and the sign of the data. 
For example, the middle-order of bit-slice can borrow $1_2$ from lower order of bit-slice and lend $1000_2$ to the higher order of bit-slice. 
On the other hand, the MSB bit-slice only borrows $1_2$ and the LSB bit-slice only lends $1000_2$.
Each order of encoded bit-slice is collected to the sub-word (16b) register. 
After collecting the four 4-bit signed bit-slices, they are sent to the RLE unit and compressed if they are all zeros.
Then, only non-zero sub-word data are transferred to the MPU cores.
As a result, the number of data transactions are significantly decreased by zero compression whereas even sign bit is added in each bit-slice during the SBR.

\begin{figure*}[!t]
\includegraphics[width=\textwidth]{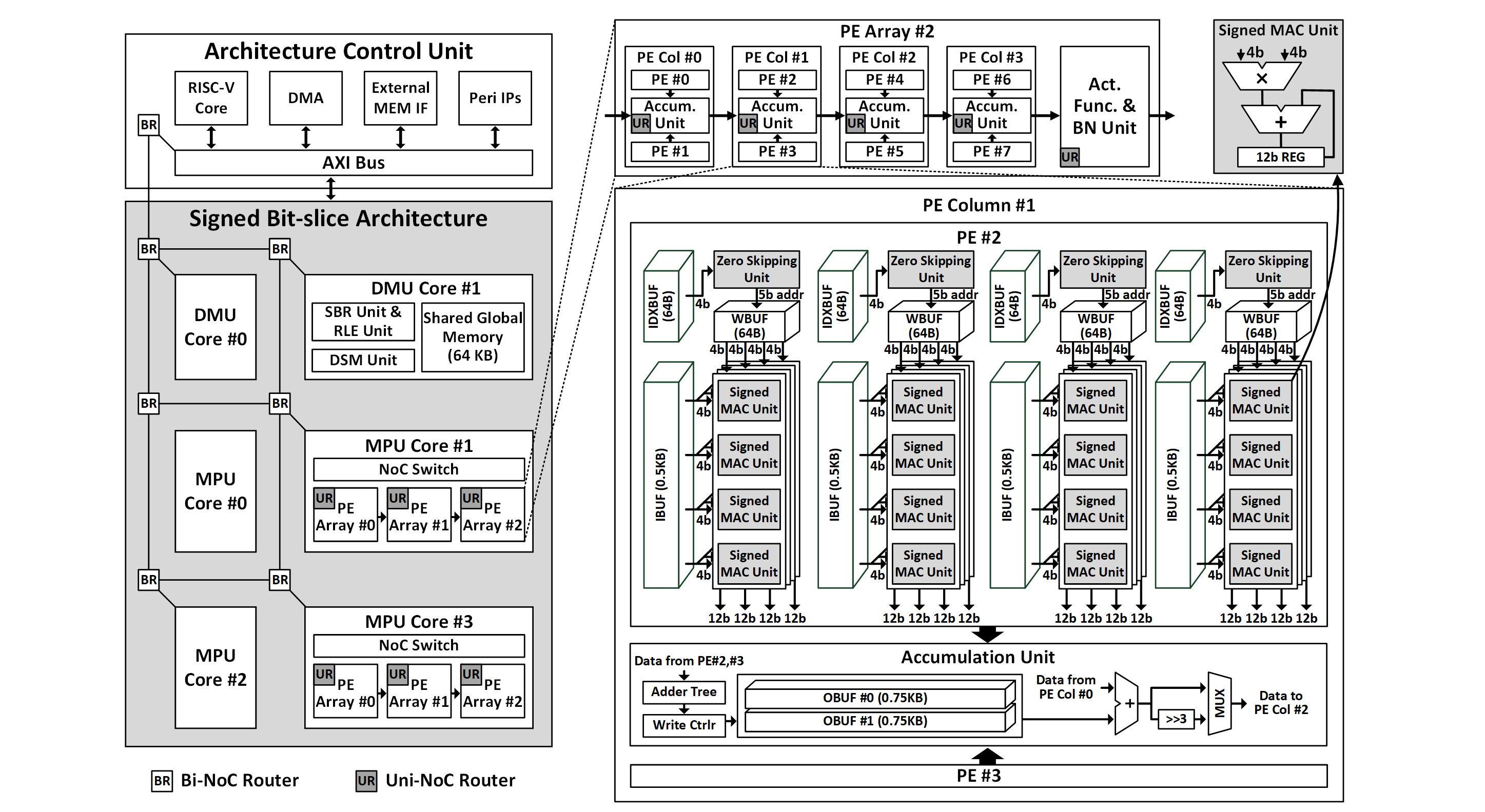} 
\caption{The overall architecture of the signed bit-slice architecture and its control unit.}
\label{fig:fig6}
\end{figure*}

\subsection{Signed MAC Unit}
Conventional bit-slice architecture integrates inefficient MAC units to compute both signed and unsigned bit-slices.
It requires a sign extension unit, an enlarged bit-width of a multiplier, and enlarged bit-width of an output accumulation register in the MAC unit.
On the other hand, since the SBR produces only signed bit-slices, the signed bit-slice architecture directly computes the signed bit-slices using the signed MAC units without sign extension.
As a result, the signed MAC unit can achieve high efficiency by reducing the bit-width of a multiplier and output accumulation register compared to the conventional MAC unit in the bit-slice architecture. 
For example, the previous work~\cite{hnpu} uses a 5b$\times$5b MAC unit with sign extension to compute 4-bit, 8-bit, 12-bit, and 16-bit precision data with its best MAC efficiency.
On the other hand, the 5b$\times$5b signed MAC unit can support 5-bit, 9-bit, 13-bit, and 17-bit precision data which shows a higher MAC efficiency than the previous work.
The other work~\cite{bitfusion} uses a 3b$\times$3b MAC unit for 2-bit, 4-bit, 6-bit, and 8-bit precision data while the 3b$\times$3b signed MAC unit supports 3-bit, 5-bit, 7-bit, and 9-bit precision.
In this paper, the signed bit-slice architecture integrates 4b$\times$4b signed MAC units to support 4-bit, 7-bit, 10-bit, and 13-bit precision data that the previous bit-slice architecture has to adopt the 5b$\times$5b MAC units to support those precision.
As a result, the signed bit-slice architecture saves 21.9\% of energy consumption of the MAC unit at 7-bit precision of DNNs compared to the conventional bit-slice architecture.

\subsection{Zero Skipping Unit for Input and Output Skipping}
Fig.~\ref{fig:fig6} describes the data path of the PE unit with the zero skipping unit.
The SBR generates a large number of 4-bit zero bit-slices even in a non-ReLU activation function, and a large number of computations can be removed.
To minimize the overheads of a fine-grained zero bit-slice skipping unit, the signed bit-slice architecture processes the four spatially adjacent 4-bit input bit-slices as sub-word data, and it skips the zero sub-word data.
The PE fetches the non-zero sub-word data from the input buffer (IBUF) and its RLE index from the index buffer (IDXBUF).
The IBUF broadcasts the input sub-word data to the four MAC arrays, and the MAC array allocates the sub-word data to the four signed MAC units by splitting it to the four 4-bit bit-slices. 
At the same time, the MAC array loads the corresponding weight bit-slice by calculating the next address of the weight buffer (WBUF) using the RLE index at the zero skipping unit.
The weight bit-slice is shared with the four signed MAC units of the MAC array generating four spatially adjacent output partial sums.
The four MAC arrays produce four different channels of output partial sums by loading four different output channels of weight bit-slices from the WBUF.
Therefore, it can skip the 16 numbers of multiplications with the zero input sub-word data.
The signed MAC unit has a 12-bit accumulation register, and accumulates the output partial sums of the subsequent input channels.
Four columns of the MAC array also compute the different input channels, so their outputs are added up at the accumulation unit after finishing the channel accumulation at the signed MAC units.

Finish time of each column of the MAC array is different during zero skipping. 
It makes the columns of MAC arrays being stalled until all of columns of MAC arrays finish the computation of allocated input channels.
To alleviate the utilization of the MAC array, the accumulation unit latches the outputs of the MAC arrays temporally to the registers.
Therefore, early finished MAC array transfers the output partial sums to the accumulation unit, and it proceeds the convolution of the next spatial input data.

In a large scale max pooling layer, a large number of convolution operations can be removed by exploiting the output sparsity. 
The signed bit-slice architecture computes the output partial sums of high order of bit slices in advance for the output speculation.
It skips the computation of remaining low order of bit-slices if they are predicted as non-maximal candidates, and only maximal candidates are completed.
For example, in a speculation of the max pooling operation, the size of values are predicted by 4-bit MSB slice convolution of inputs ($I_M$) and weights ($W_M$).
If they are predicted as non-maximal outputs, remaining low order of bit-slice computations, $I_M \times W_L$, $I_L \times W_M$, and $I_L \times W_L$, are skipped.
To support the output skipping at the zero skipping unit, corresponding input channels of input data are set to zeros, and they are skipped by input skipping.
Therefore, maximal outputs are masked with a binary map and the RLE unit loads the binary map and regenerates the non-zero in-out sub-word data.
Finally, the non-zero sub-word data and their index are transferred to the PEs, where zero skipping operation is performed.
To minimize the complexity of the zero skipping unit, successive four output channels of non-maximal outputs are skipped.

\subsection{Hybrid Zero Skipping and Compression with Dynamic Sparsity Monitoring}
The sparsity of high order of bit-slices is high by using a SBR, but the sparsity of low order is usually low especially in a non-ReLU activation function. 
Among the convolution of bit-slices, $I_M \times W_M$ and $I_M \times W_L$ shows a high performance due to high sparse $I_M$.
However, dense $I_L$ degrades the performance of $I_L \times W_M$ and $I_L \times W_L$, which decreases the overall performance.
Therefore, the weight bit-slices are exploited for skipping instead of input bit-slices if the sparsity of them is higher than inputs.
Moreover, to increase the efficiency of $I_L \times W_L$ when both $I_L$ and $W_L$ are dense, The signed bit-slice architecture disables the zero skipping units and IDXBUFs during computation of dense bit-slices to reduce the dynamic power.
Similarly, the compression of the dense bit-slices deteriorates the compression ratio due to relatively high bit-width of non-zero index.
Therefore, the compression is not applied to them but still maintains high compression ratio by high order of sparse signed bit-slices.

To support the weight skipping at the zero skipping unit, four adjacent output channels of weight bit-slices become the sub-word data, and this sub-word data is allocated to the IBUF and corresponding four adjacent spatial input bit-slices are fetched to the WBUF.
Then, the accumulation unit rearranges the computation results before storing to the OBUF to match the same pattern of the input sub-word.
For the weight and output skipping, the four adjacent output channels of weights corresponding to non-maximal outputs are encoded to zeros.
Consequently, the signed bit-slice architecture supports input, weight, and output skipping without changes of data-path.

The DSM unit monitors the sparsity of the data on a run-time and makes a decision to exploit the sparsity exploitation.
Decision is done during loading the input and weight data from the external memory and storing the data to the global memory after generating the final convolution outputs.
As a result, it shows a high performance of zero skipping by selecting the higher sparse data and high compression ratio by deciding the compression operation only on high sparse data.

\begin{figure} [!t]
\includegraphics[width=\columnwidth]{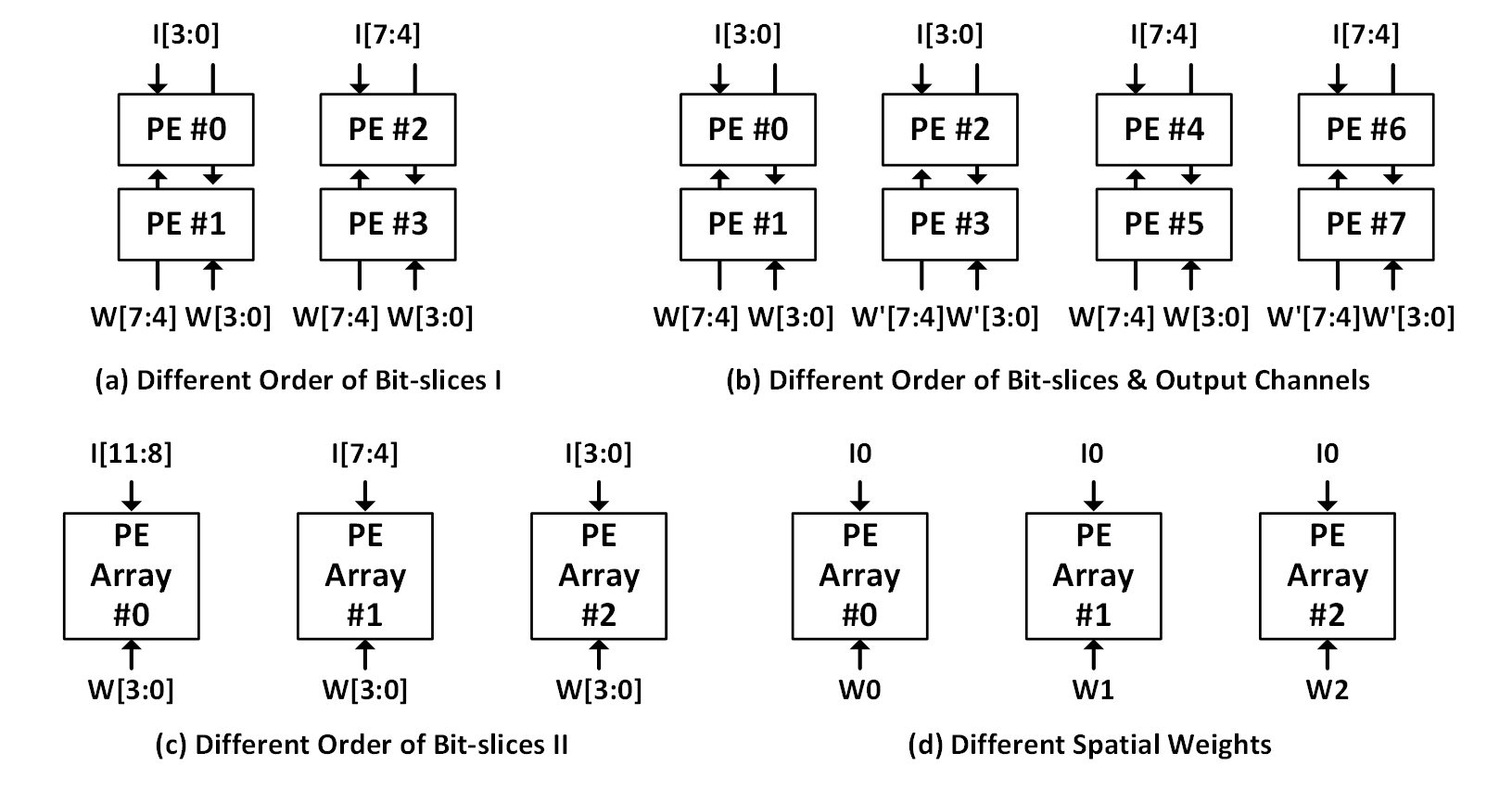} 
\centering
\caption{Workload allocation with exploitation of data reusability.}
\label{fig:fig7}
\end{figure}

\begin{figure} [!t]
\includegraphics[width=\columnwidth]{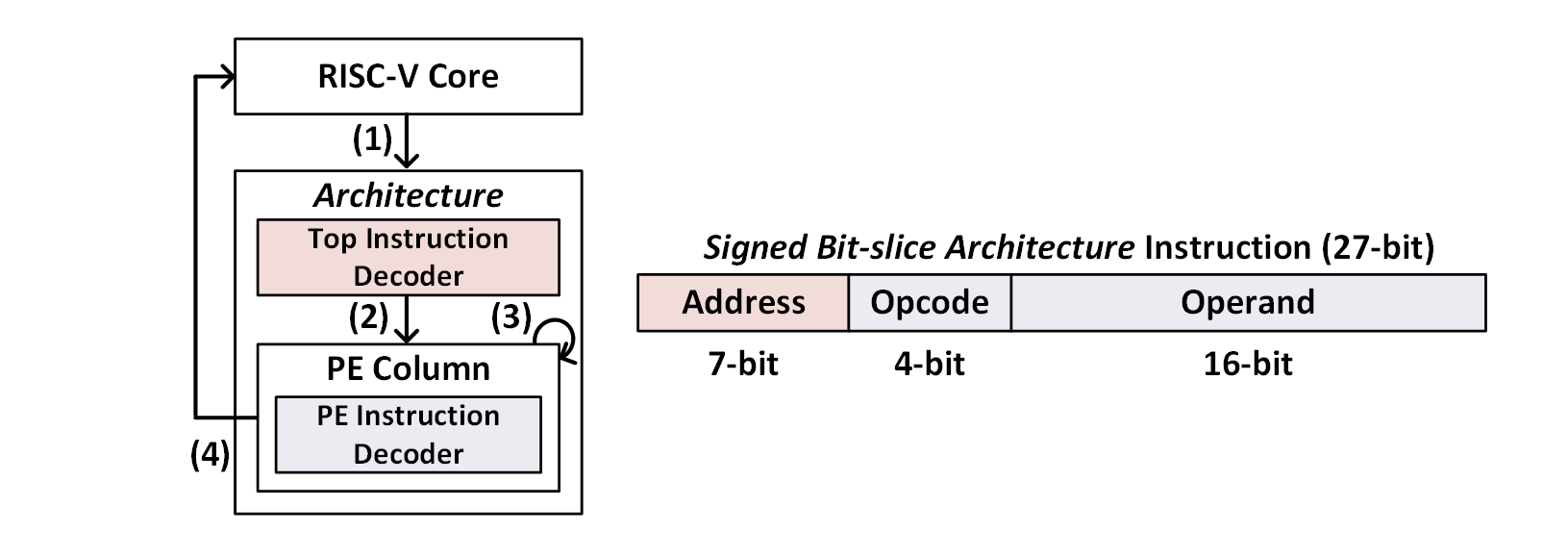} 
\centering
\caption{The signed bit-slice architecture ISA and hierarchical instruction decoder.}
\label{fig:fig8}
\end{figure}

\subsection{Heterogeneous Network-on-chip}
Heterogeneous NoC is adopted for efficient DNN workload allocation as shown in Fig.~\ref{fig:fig6}.
For the input, weight, and output data transmission, the bi-directional Bi-NoC is used.
The data management unit (DMU) core provides the input and weight data to the matrix processing unit (MPU) cores, and the MPU cores transfer the convolution outputs to the DMU core through the Bi-NoC. 
The Bi-NoC also flexibly transfers weight data to IBUF and input data to WBUF for hybrid skipping.
After receiving the data through the router, the NoC switch unicasts, multicasts, and broadcasts the data to the PE arrays with the data reusability.
For example, the input bit-slice I[3:0] is multicasted to the PE\#0 and PE\#1 and weight bit-slice W[3:0] is multicasted to the PE\#1 and PE\#3 as shown in Fig.~\ref{fig:fig7}(a). Then, by allocating the different output channels of weights, the input bit-slice I[3:0] is reused to four PEs in Fig.~\ref{fig:fig7}(b).
The weight bit-slice W[3:0] is broadcasted to the three PE arrays to compute the different orders of input bit-slice in Fig.~\ref{fig:fig7}(c).
The input data is shared with three PE arrays by unicasting the different spatial weights of 3$\times$3 weights in Fig.~\ref{fig:fig7}(d).
As a result, the various combinations of workload allocation is exploited in the signed bit-slice architecture for the data reusability through the Bi-NoC.

Output partial sum requires a large bit-width compared to the input and weight data, and its bit-width is also enlarged after applying a arithmetic shift operation to accumulate the other order of partial sums.
To minimize the transmission bandwidth of the partial sums, each accumulation unit applies the right arithmetic shift by 3 to the partial sums before passing to the subsequent accumulation unit.
Then, Uni-NoC transfers the reduced bit-width of partial sums and the subsequent accumulation unit accumulates them to obtain the final convolution outputs.
As a result, the right arithmetic shift unit reduces the bandwidth of Uni-NoC by 40\% compared to the previous bit-slice architecture~\cite{hnpu}.
Uni-NoC connects two adjacent accumulation units and routes the data from right to left.
Then, the signed bit-slice architecture allocates the low order of bit-slices to the left side of the PE and high order of bit-slices to the right side of the PE to accumulate partial sums across the PEs with Uni-NoC.
If the two adjacent accumulation units generate the same order of the output partial sums, the partial sums are just passed to the next accumulation unit without a bit-shift operation.
As a result, Uni-NoC transfer the partial sums with minimum bandwidth.

\subsection{Instruction Set Architecture}
The instruction of the CPU (e.g. RISC-V ISA) is not compatible to the DNN execution which consists of a large number of iterative MAC operations. 
Since the CPU is busy to control other hardware units (e.g. GPU, I/O) in a mobile SoC, the NPU has to be controlled with minimum involvement of the CPU. 
Therefore, the specialized ISA and hierarchical instruction decoder for the signed bit-slice architecture is provided as shown in Fig.~\ref{fig:fig8}.
The signed bit-slice architecture fetches the instructions from the RISC-V core (1), and the top instruction decoder reads the high 7-bit of instructions and sends the opcode and operand to the target address of DMU cores or MPU cores (2).
Then, the instruction decoder inside the target unit decodes the 4-bit opcode and 16-bit operand of the remain instruction, and it configures and activates the cores (3).
The opcode and operand specify the configuration of MPU core, accumulation unit, DMU core, and reset and run signal of the core and unit.
Before activating the PEs, the configuration instructions setup the PEs by specifying the size of the workloads such as the information of width, height, input channel, and output channels of the tiled data.
Then, the run instruction triggers the PEs to perform the convolution operation. 
The PE generates the address of the IBUF and WBUF itself to load the data without involvement of the CPU, and computes the convolution of tiled data.
After reaching the end address of the target workload, the signed bit-slice architecture fetches the run instruction again without setup of the PEs for the next tiled data if the configuration is the same as before (4).
As a result, the specialized ISA reduces the number of instruction fetch operations to minimize the involvement of the CPU.

\section{Evaluation}
\subsection{Benchmarks}
The signed bit-slice architecture is evaluated on the various 2D and 3D DNNs: YoloV3~\cite{yolov3}, Monodepth2~\cite{monodepth2}, VoteNet~\cite{votenet}, and DGCNN~\cite{dgcnn}.
YoloV3 is a popular 2D object detection application which is a dense DNN with the leaky-ReLU activation function. 
It shows a 29.2$\%$ of input sparsity in 7-bit precision.
Monodepth2 is a monocular depth estimation application which consists of a encoder network and a decoder network.
Although the encoder network requires a low bit-precision (7-bit) with a high input sparsity (57.3$\%$) due to ReLU activation function, the decoder network requires 10-bit precision of inputs and 7-bit precision of weights without any loss of accuracy.
It has a low input sparsity (17.5$\%$) because of the ELU activation function.
VoteNet is a 3D object detection application which is based on a PointNet++~\cite{pointnet++}.
It has 7-bit precision of inputs and weights, and it shows a $46.2\%$ of input sparsity with a 64-to-1, a 32-to-1, and three 16-to-1 max pooling layers.
DGCNN is a graph neural network for 3D vision tasks and is composed of leaky-ReLU activation functions and four 40-to-1 max pooling layers.
It shows a 17.3$\%$ of input sparsity with 7-bit precision.
Performance of VoteNet and DGCNN excludes the point processing such as neighbor search and point sampling.

\subsection{Evaluation Methodology}
\begin{figure} [!t]
\includegraphics[width=\columnwidth]{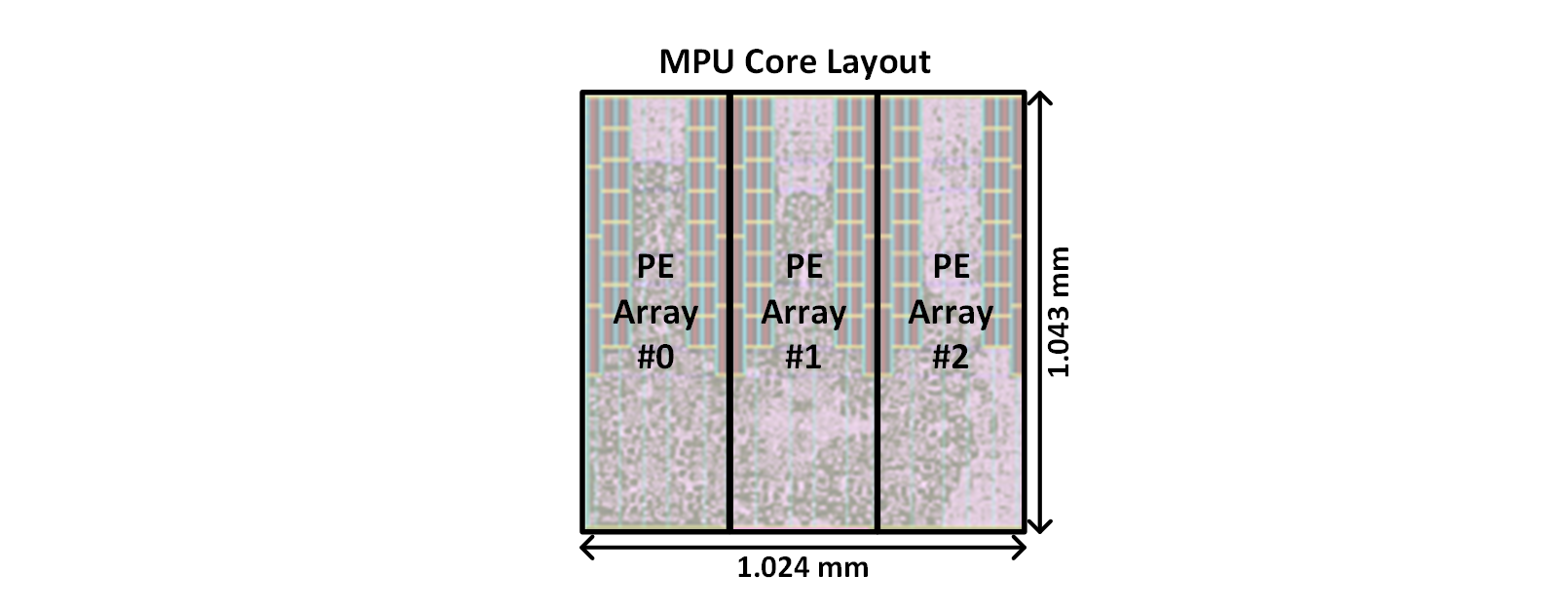} 
\centering
\caption{The layout of the MPU core.}
\label{fig:fig13}
\end{figure}

The signed bit-slice architecture is designed in RTL verilog and evaluated through RTL simulations to count exact cycles during the DNNs execution.
It is synthesized in Samsung 28 nm CMOS technology using Synopsys Design Compiler and designed with Synopsys IC Compiler II for the place and route.
Fig.~\ref{fig:fig13} shows the post-layout of the MPU core in the signed bit-slice architecture.
The three PE arrays are aligned horizontally in the $1.024 mm$ $\times$ $1.043 mm$ sized MPU core.
In the layout, the top of power metal layer in the MPU core is IA layer, and and top of signal metal layer is M7 layer.
The power consumption is measured with the Synopsys PrimeTime after the place and route.

The RISC-V core (RV64IMAFDC) in the architecture control unit is used for the overall control of the signed bit-slice architecture, and also compiles the DNN benchmarks to the specialized ISA.
Then, the RISC-V core loads the source code of the DNNs execution and allocates the workloads to the MPU cores through Bi-NoC.
Off-chip memory is modeled using Cypress Semiconductor's HyperRAM for a external IoT DRAM.
Then, the architecture control unit integrates the HyperRAM interface to access the HyperRAMs.

\subsection{Comparison with Previous Accelerators}
\begin{figure} [!t]
\includegraphics[width=\columnwidth]{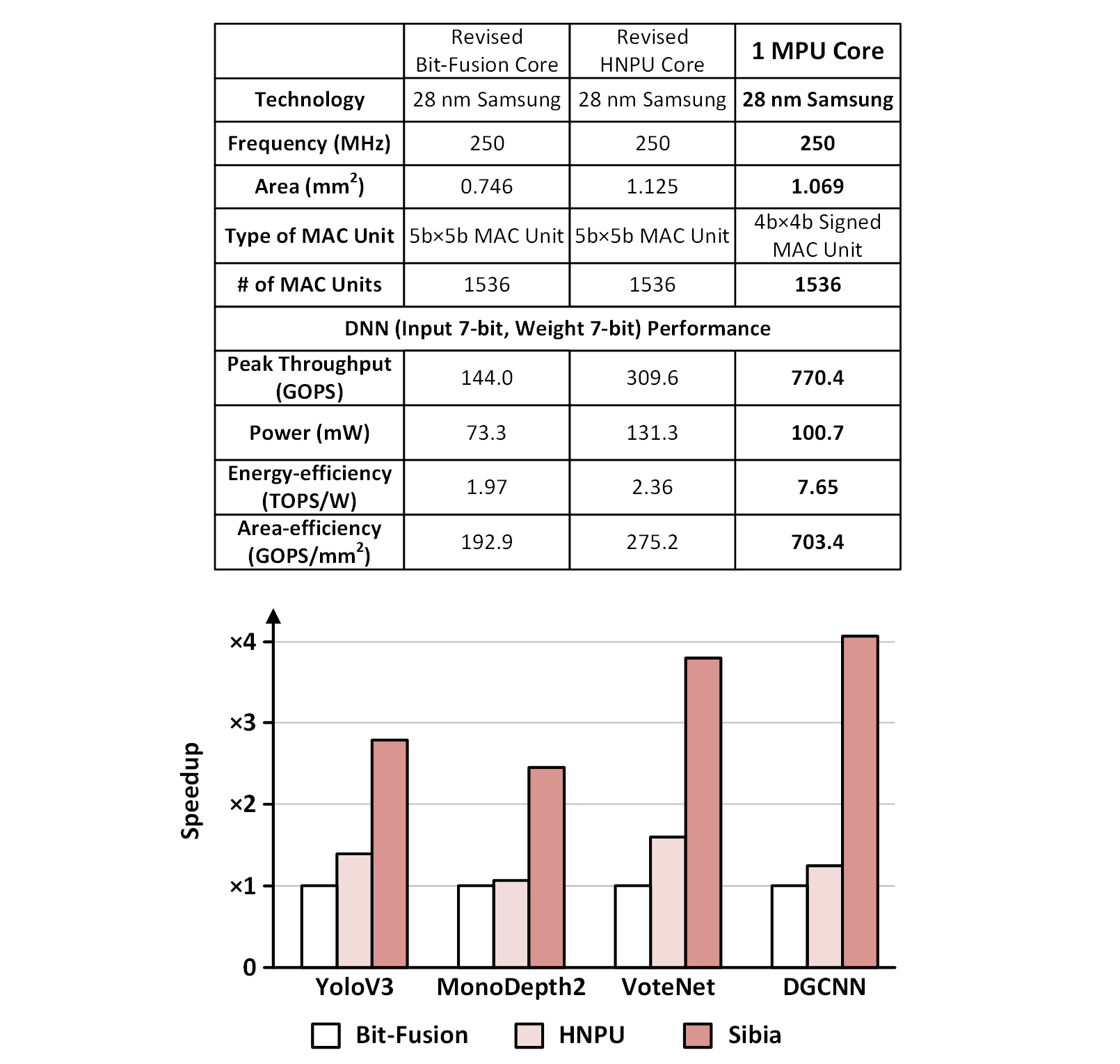} 
\centering
\caption{Speedup comparison with previous accelerators.}
\label{fig:fig10}
\end{figure}

The signed bit-slice architecture is compared with the previous bit-slice accelerator, Bit-fusion~\cite{bitfusion}, and bit-slice skipping accelerator, HNPU~\cite{hnpu}.
For a fair comparison, this paper revised a Bit-fusion and HNPU to match the same number of MAC units, the same technology node, and the same clock frequency.
Fig.~\ref{fig:fig10} illustrates the comparison of three accelerators.
The 1.069 $mm^{2}$ MPU core in the signed bit-slice architecture achives 770.4 GOPS of peak throughput at 7-bit DNN performance while consuming 100.7 $mW$ of power.
Although area and power consumption of the MPU core is $\times1.43$ larger and $\times1.37$ higher than Bit-fusion respectively, the peak throughput of the MPU core achieves $\times5.35$ higher.
Then, the area-efficiency and energy-efficiency of the MPU core is $\times3.65$ and $\times3.88$ higher than Bit-fusion, respectively.
In the comparison with HNPU, the peak throughput of the MPU core is $\times2.49$ higher while the area and power consumption is $95.0\%$ and $76.7\%$ of HNPU, respectively.
Therefore, the area-efficiency and energy-efficiency of the MPU core is $\times2.56$ higher $\times3.24$ higher than HNPU.

In the DNN benchmarks, zero bit-slice skipping of HNPU shows a small speedup in dense DNNs, $\times1.35$, $\times1.08$, and $\times1.24$ speedup in YoloV3, Monodepth2, and DGCNN respectively.
On the other hand, all of the inference speed of the signed bit-slice architecture outperforms Bit-fusion and HNPU. 
It shows $\times2.79$ speedup in YoloV3 and $\times2.48$ speedup in Monodepth2 by using hybrid skipping method even in the dense DNNs.
It also achieves $\times3.73$ and $\times4.11$ faster than the baseline in VoteNet and DGCNN respectively by using in-out skipping method.

\begin{figure} [!t]
\includegraphics[width=\columnwidth]{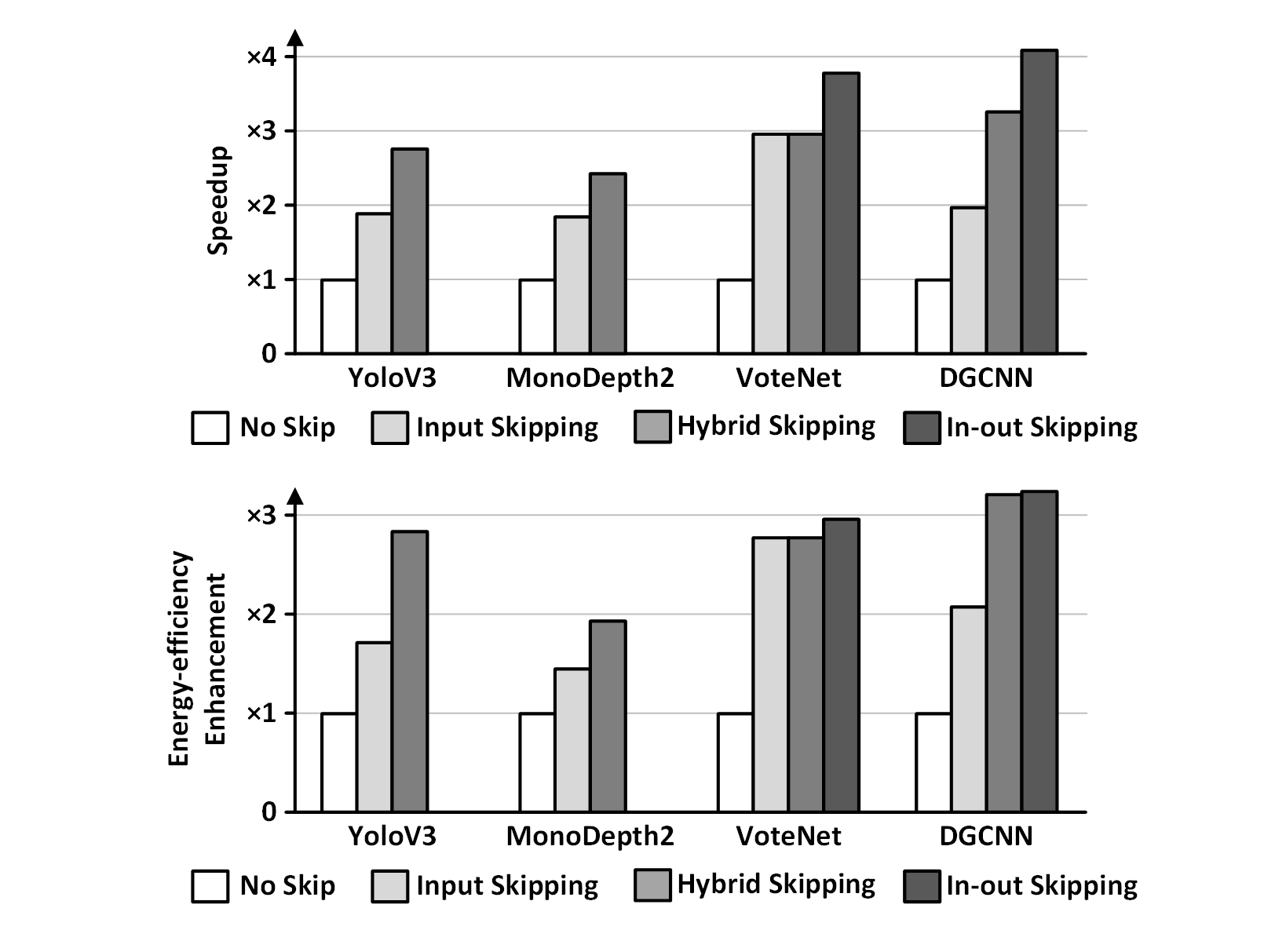} 
\centering
\caption{Speedup and energy-efficiency improvement with various skipping modes}
\label{fig:fig11}
\end{figure}

\begin{figure} [!t]
\includegraphics[width=\columnwidth]{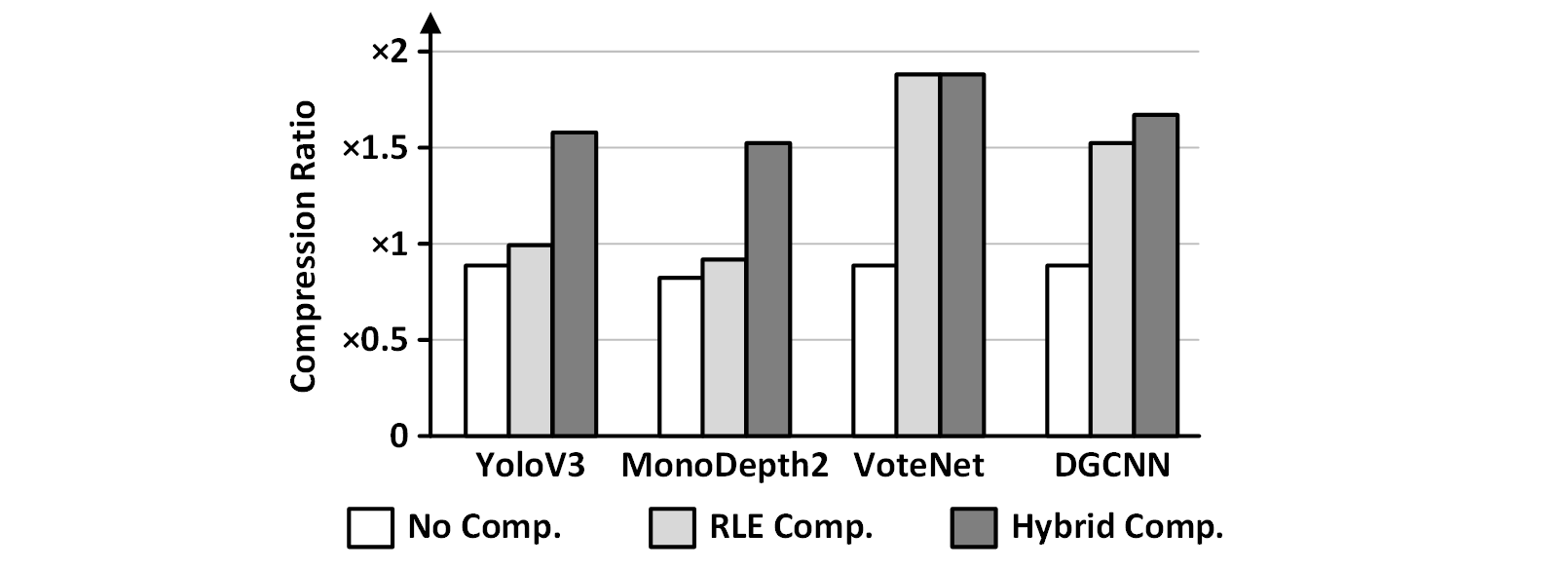} 
\centering
\caption{Compression ratio with various compression modes}
\label{fig:fig14}
\end{figure}

\subsection{Performance Improvement}
Fig.~\ref{fig:fig11} shows the performance improvement by no skip mode, input skipping mode, hybrid skipping mode, and in-out skipping mode in the signed bit-slice architecture.
Input skipping increases the inference speed by $\times1.88$, $\times1.86$, $\times2.94$, and $\times2.15$ higher in YoloV3, Monodepth2, VoteNet, and DGCNN, respectively.
By exploiting the sparse weight signed bit-slices, hybrid skipping boosts up the throughput by $\times2.79$, $\times2.48$, $\times2.94$, and $\times3.28$ higher in YoloV3, Monodepth2, VoteNet, and DGCNN, respectively.
Since VoteNet presents sparse inputs data by ReLU activation function, its throughput enhancement by input skipping is the highest among the benchmarks, and its throughput enhancement by hybrid skipping is the same with input skipping because its sparsity of high order of weight signed bit-slice is lower than the sparsity of low order of inputs at the most of the convolution layers.
Despite of the dense inputs and weights data in YoloV3, Monodepth2, and DGCNN, throughput enhancement of them are close to and even higher than the sparse VoteNet by using hybrid skipping.
In-out skipping at the large scale of max pooling layers additionally increases throughput, achieving $\times3.73$ in VoteNet and $\times4.11$ in DGCNN.

\begin{figure*}[!t]
\includegraphics[width=\textwidth]{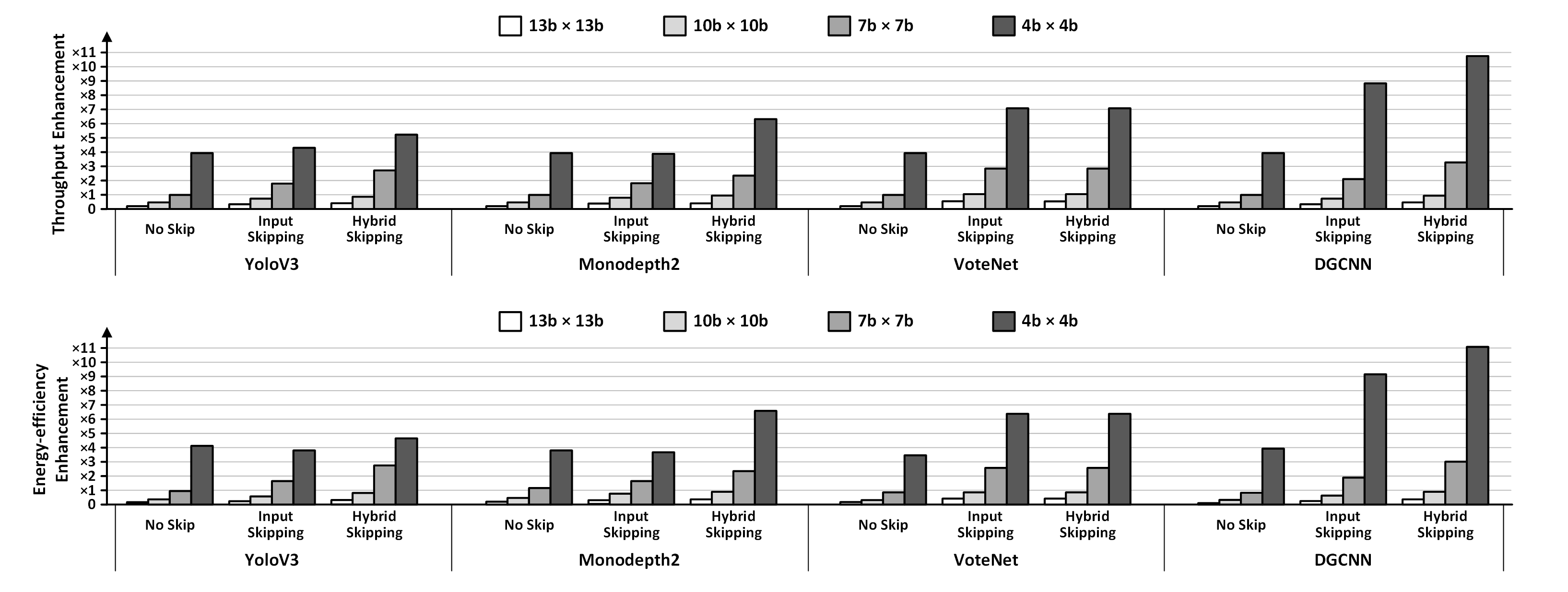} 
\caption{Throughput enhancement with various bit-precision of DNN benchmarks}
\label{fig:fig17}
\end{figure*}

\begin{figure} [t]
\includegraphics[width=\columnwidth]{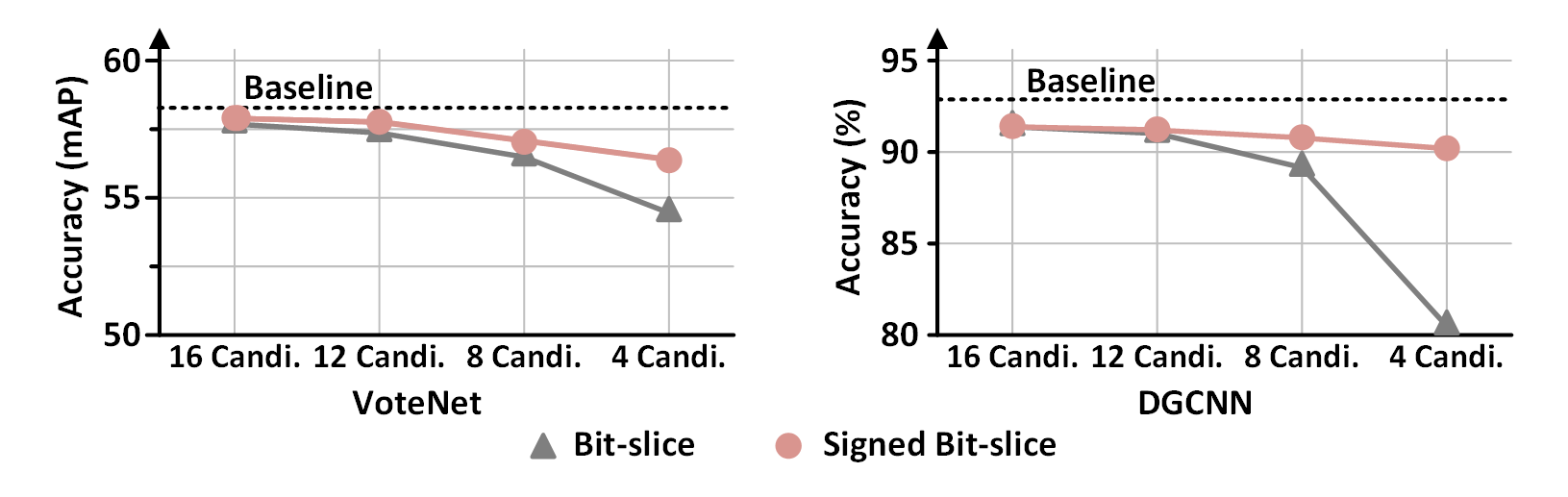} 
\centering
\caption{DNN accuracy after output speculation with the different number of candidates.}
\label{fig:fig15}
\end{figure}

\begin{figure} [!t]
\includegraphics[width=\columnwidth]{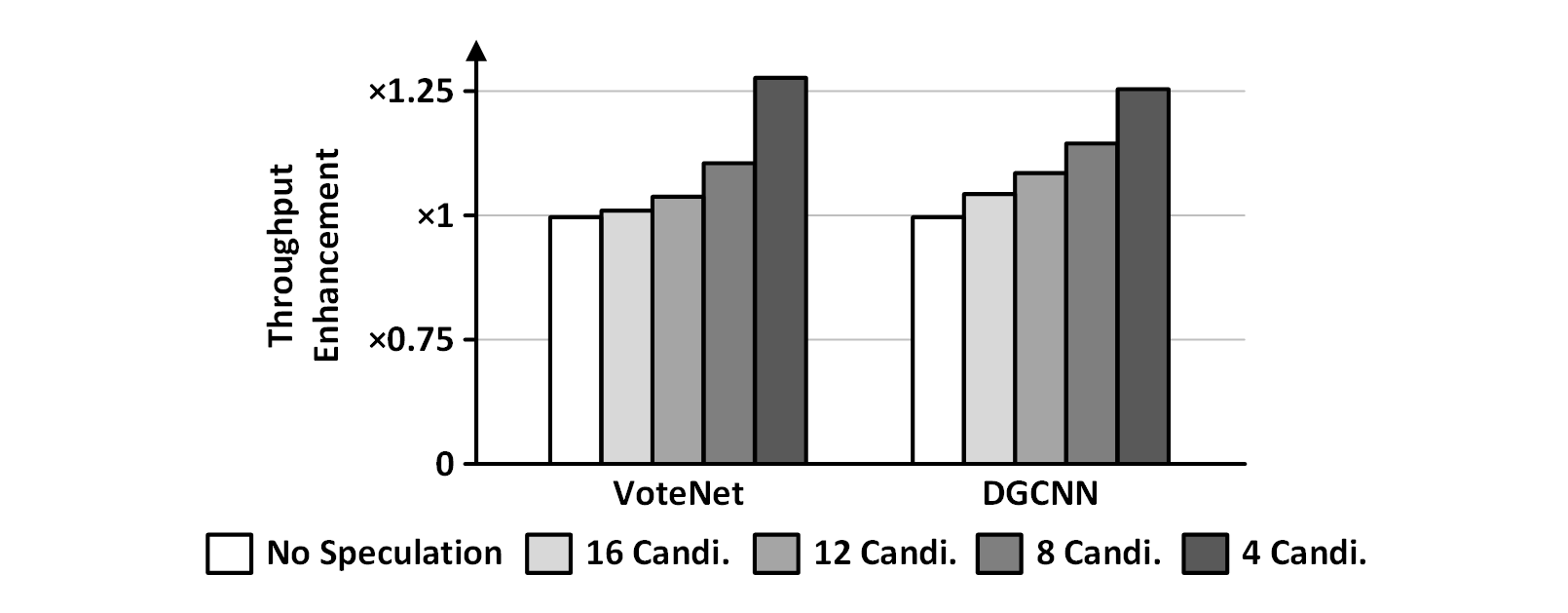} 
\centering
\caption{Throughput enhancement and energy-efficiency enhancement of in-out skipping compared to hybrid skipping (no speculation) with the different numbers of speculation candidates}
\label{fig:fig16}
\end{figure}

The energy-efficiency increases by using input skipping and in-out skipping.
Similar to the throughput enhancement, energy-efficiency of VoteNet shows the highest enhancement ($\times2.80$) in input skipping among the benchmarks.
Then, hybrid skipping increases the energy-efficiency of YoloV3, MonoDepth2, and DGCNN by $\times2.82$, $\times1.90$, and $\times3.33$ higher than the baseline repectively, which is close to energy-efficiency of sparse VoteNet.
In-out skipping achieves slightly higher energy-efficiency than the hybrid skipping because additional power consumption is occurred during output speculation.
Finally, the energy-efficiency enhancement of DGCNN measures $\times3.36$ which is the most highest among the benchmarks.

Fig.~\ref{fig:fig14} shows the compression ratio of input signed bit-slices in each benchmark.
Since the SBR adds the 1-bit sign bit to each bit-slice, the size of the raw signed bit-slices (No compression in Fig.~\ref{fig:fig14}) becomes bigger than the baseline.
Then, the SBR generates a large number of zero bit-slices, so VoteNet and DGCNN shows $\times$1.81 and $\times$1.54 compression ratio respectively by using RLE compression.
However, the low order of signed bit-slices easily become dense data which deteriorates the compression ratio, resulting low compression ratio at YoloV3 and MonoDepth2.
Therefore, the dense data are not compressed (hybrid compression in Fig.~\ref{fig:fig14}), and the compression ratio of YoloV3 and MonoDepth2 achieves $\times$1.57 and $\times$1.54 respectively.
In VoteNet, low order of signed bit-slices are sparse enough so that hybrid compression is not required.
Consequently, encoding to the signed bit-slices achieves high compression ratio even in the dense DNNs.

Fig.~\ref{fig:fig17} illustrates the throughput enhancement by bit-precision of data.
All of the DNN benchmarks quantized the inputs and weights data to 4-bit, 7-bit, 10-bit, and 13-bit.
The baseline ($\times1$) is DNN computation of $7b\times7b$ bit-precision without any skipping.
As increasing the bit-precision of DNNs, the throughput decreases quadratically because the MAC units have to compute the increased number of bit-slices of both inputs and weights.
Therefore, the throughput enhancement without skipping presents $\times\sfrac{1}{16}$ in $13b\times13b$, $\times\sfrac{1}{4}$ in $10b\times10b$, and $\times4$ in $4b\times4b$ compared to $7b\times7b$ bit-precision, respectively.
By using input skipping, decreasing ratio of throughput is close to linear.
For example, the throughput decrease ratio in Monodepth2 is $\times0.48$ from $10b\times10b$ to $13b\times13b$ bit-precision, $\times0.44$ from $7b\times7b$ to $10b\times10b$ bit-precision, and $\times0.47$ from $4b\times4b$ to $7b\times7b$ bit-precision.
Hybrid skipping improves the throughput much higher, and the inference speed of $13b\times13b$ by hybrid skipping is similar to and even faster than the $10b\times10b$ bit-precision without skipping.
Although only input or weight skipping on $7b\times7b$ bit-precision cannot surpass the speed of original $4b\times4b$ bit-precision, output skipping finally achieves much faster speed in DGCNN as shown in Fig.~\ref{fig:fig11}.
The baseline ($\times1$) of energy-efficiency enhancement is DNN computation of $7b\times7b$ bit-precision in YoloV3 without skipping.
Energy-efficiency enhancement shows a similar trend of the throughput enhancement.
However, the energy-efficiency of input skipping in dense Yolov3 and Monodepth2 are much lower than their no skip mode in $4b\times4b$ bit-precision because they consume more power by activating the zero skipping unit and IDXBUF with low throughput enhancement.

Fig.~\ref{fig:fig15} shows DNN accuracy by output speculation.
For the output speculation of VoteNet, pre-computation of $I_M \times W_M$ at the 64-to-1 and 32-to-1 max pooling layers and $I_M \times W_M$ + $I_L \times W_M$ at the three 16-to-1 max pooling layers is used for the minimum accuracy loss.
Then, DGCNN requires $I_M \times W_M$ + $I_L \times W_M$ for all of max pooling layers.
With the small number of candidates, DNN accuracy is rapidly degraded in conventional bit-slice due to its asymmetric numbers.
On the other hand, signed bit-slice increases the success rate of speculation and achieves the minimum accuracy loss (${\sim}2\%$) in both DNNs while using 4 candidates for the speculation.
Fig.~\ref{fig:fig16} describes the throughput enhancement of in-out skipping by the number of the candidates for output speculation.
As increasing the number of candidates, the throughput of the signed bit-slice architecture increases in exponential.
Then, the throughput enhancement of VoteNet and DGCNN achieves $\times1.27$ and $\times1.25$ higher than hybrid skipping (no speculation in Fig.~\ref{fig:fig16}) at 4 candidates, respectively.

\subsection{Area and Power Breakdown}
\begin{figure} [!t]
\includegraphics[width=\columnwidth]{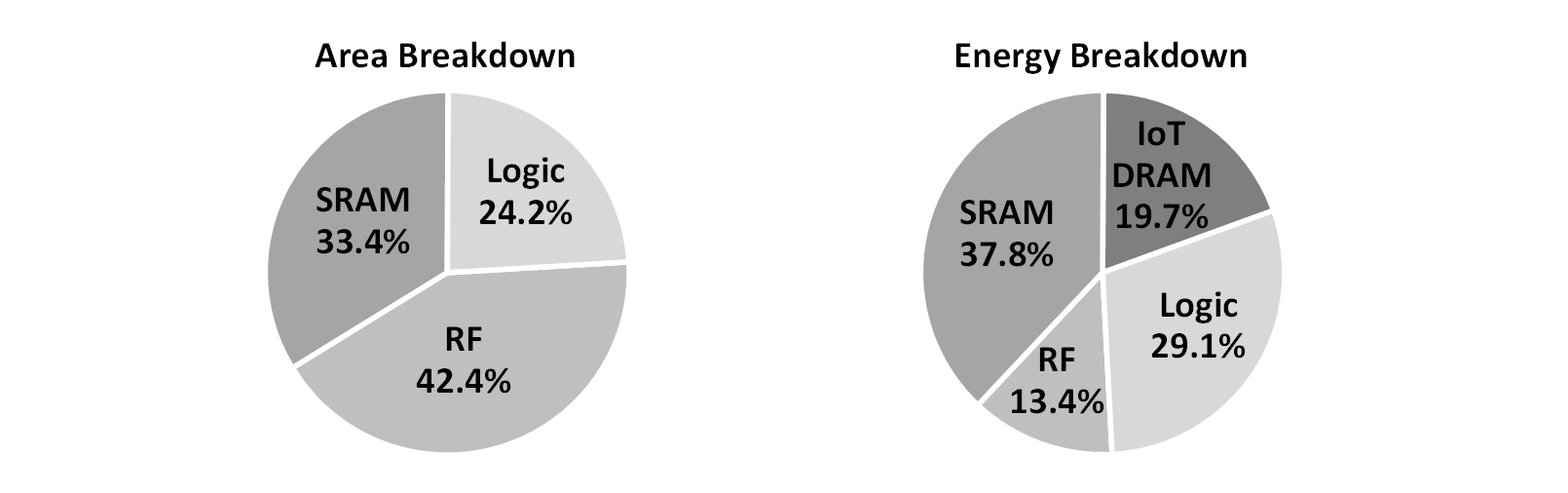} 
\centering
\caption{Area and energy breakdown of the signed bit-slice architecture}
\label{fig:fig12}
\end{figure}

Fig.~\ref{fig:fig12} illustrates the area and energy breakdown of the signed bit-slice architecture.
The area breakdown is measured by Synopsys Design Compiler after logic synthesis.
The register file (RF) and on-chip SRAM takes up 42.4\% and 33.4\% of the total area, respectively.
Then, control and compute logic accounts for 24.2\% of the total area.

In the energy breakdown, the on-chip SRAM, RF, and logic takes up the 37.8\%, 13.4\%, and 29.1\% of the overall energy consumption, respectively.
The IoT DRAM is modeled by Cypress Semiconductor's HyperRAM, and its energy consumption is estimated by counting the read and write time of the DRAMs.
Then, the DRAM accounts for 19.7\% of the overall energy consumption.

%%%%%%% -- PAPER CONTENT ENDS -- %%%%%%%%
\section{Conclusion}
The paper presents the energy-efficient signed bit-slice architecture for dense DNN acceleration on mobile platforms.
Unlike a conventional bit-slice decomposition method, the signed bit-slice representation adds the sign bit in each unsigned bit-slice.
As a result, it generates a lot of zero bit-slices and balances 2's complement data between positive and negative values.
With these advantages, the signed bit-slice architecture increases the hardware performance by using zero input bit-slice skipping with the efficient signed MAC units.
The signed bit-slice architecture is also compatible with output skipping using the zero input bit-slice skipping unit by masking the speculated input data to zeros.
Moreover, the signed bit-slice architecture supports zero weight singed bit-slice skipping by encoding the weight data instead of input data and feeding it to the IBUF and the IDXBUF.
Therefore, the signed bit-slice architecture can skip more sparse data among input and weight, which boosts up the throughput and energy-efficiency.
The heterogeneous NoC flexibly transfers the input and weight data by considering the data reusability and reduces the transmission bandwidth of the output partial sum data.
A specialized ISA of the signed bit-slice architecture and the hierarchical instruction decoder optimizes the DNN workloads and minimizes the involvement of the CPU.
As a result, the signed bit-slice architecture outperforms the previous bit-slice accelerators in dense 2-D and 3-D DNN benchmarks.

%%%%%%%%% -- BIB STYLE AND FILE -- %%%%%%%%
\bibliographystyle{IEEEtranS}
\bibliography{refs}
%%%%%%%%%%%%%%%%%%%%%%%%%%%%%%%%%%%%

\end{document}